# Effects of li doping on superconducting properties of citrate-gel prepared $Y_{1-x}Li_xBa_2Cu_3O_{7-\delta}$ Compound

Mahshid Amiri–Moghadam[1] · Hassan Gashmard[1] · Seyed Sajjad Hosseini[1,2] · Hamideh Shakeripour[1] · Hadi Salamati[1]

**Abstract**

The $Y_{1-x}Li_xBa_2Cu_3O_{7-\delta}$ polycrystalline bulk superconductors doped with Li substituting at the Y site at different concentrations ($x = 0, 0.01, 0.02, 0.1$) were prepared using the citrate-gel method to study the effects of doping on the superconducting temperature and critical current density. The question was whether Li addition characterized by a high Debye frequency would have any positive effects on $T_c$. The optimum citrate-gel and heat treatment conditions were identified as those yielding samples with a maximum grain size on the order of 50 μm (up to the optimum Li doping level, $x = 0.01$). Li substitution at the Y site was verified by structural, electrical, and magnetic measurements of the produced samples, whereas X-ray diffraction (XRD) analysis revealed the formation of a pure phase with no visible impurity phases. Moreover, AC magnetic susceptibility measurements showed no increase in the superconducting transition temperature $T_c$, consistent with the predicted results obtained by machine learning method, although it was theoretically expected to increase owing to the high Debye frequency of Li. This observation is consistent with magnetic coupling models for pairing mechanism in cuprates. Finally, because of the optimum conditions of the preparation procedure, nearly identical values of the critical current density ($J_c$) were recorded for samples with different Li doping levels (up to the optimum Li doping level). It was found that improved compound preparation conditions would have a critical and extensive effect on $J_c$ enhancement, with nearly no $T_c$ suppression.

## 1 Introduction

Superconductivity is among the most important discoveries of the twentieth century [1]. High-temperature superconductors (HTS) have found numerous technological applications such as MRI systems, electromagnets, transformers, and high-speed trains [2]. Many studies have focused on $YBa_2Cu_3O_{7-\delta}$ (Y–123) cuprates and their modifications to improve their $T_c$ and $J_c$ [3–5]. The great interest in the study of Y–123 properties stems from the fact that the physical properties of Y–123 systems can be modified over a wide range of doping levels without any significant changes in the lattice structure [6, 7].

The fundamental mechanism of superconductivity of the cuprate family is not yet well-known and numerous theoretical models attempt to explain the phenomenon. The substitution of different elements in the Y–123 structure is generally meant to understand mechanism of cuprates, to improve its superconducting properties as well as its physical properties for various applications. One such property is its high critical current density, which is essential for the industrial applications of high-temperature superconductors (HTS). Since the discovery of Y-based cuprate superconductors, substitution of elements at different atomic sites, especially at the Cu site, has been widely studied to find that elements such as Li, Zn, or Ni substituted at the different cationic sites in the Y–123 structure would act as current pinning centers [8–10]. Furthermore, elements with suitable covalent radii (Co, Li, Ni, and Pd) substituted for Cu have been reported to improve the value for $J_c$ but to reduce that of $T_c$ [11]. The doping suppresses superconductivity locally and can act as pinning

✉ Hamideh Shakeripour
 hshakeri@iut.ac.ir

1 Department of Physics, Isfahan University of Technology, Isfahan 84156-83111, Iran

2 Research Institute for Nanotechnology and Advanced Materials, Isfahan University of Technology, Isfahan 84156-83111, Iran

centers resulting in an increase in critical current density. It has also been reported that flux pinning and $J_c$ enhancement due to Zn and Li doping at the Cu site ($YBa_2(Cu_{1-y}Zn_y)_3O_{6.64}$ with $y > 0.5\%$ for Zn and in $YBa_2Cu_{(100\%-x}Li_x)_{3/100\%}O_7$ with $x > 0.06$ at.% for Li) could be attributed to the formation of local magnetic moments in the proximity of non-magnetic ions such as Li [11–14]. Neutron diffraction experiments have revealed that Cu in $CuO_2$ planes is replaced with Li [15–18], thereby reducing the number of holes in the copper planes that ultimately leads to $T_c$ degradation [19]. Also, Cu substituted with Li and Ba substituted with K or Na have been reported to lead to declines in $T_c$ with the lowest $T_c$ obtained with Li substitution (in $YBa_2Cu_3$–$_xLi_xO_7$ with $x = 0$, 0.05 and 0.2) [20]. Changes in the distribution of carriers have also been claimed as another cause underlying $T_c$ reduction in Li–Cu materials. In general, Li substitution for Cu in YBCO reportedly reduces $T_c$, c-axis length, and unit cell volume [15–17, 19–21], indicating that Li actually enters the crystal structure.

From a different perspective, alkali metals with smaller ionic radii, such as Li, have been observed to cause more superconductivity suppression than the ones with larger ionic radii (i.e., Na and K). From a theoretical perspective, substitution of K gives rise to changes in the electronic band structure and density of states (DOS) that lead to declining $T_c$ [19]. It has been established that substitution of $K^+$ ions at the $Ba^{2+}$ site leads to the hole doping effect that leads to a slight increase *in* $T_c$ [22]. In a computational study, the only study in the literature about Li addition for Y site of YBCO compound, Li substitution was observed to lead to slight increases in $T_c$ at low doping levels due to the strengthened two-dimensional character of the bonds in $CuO_2$ planes giving rise [19].

Although the present work had been conceived and initiated prior to the authors' learning of these earlier findings, it had been motivated by a study of Li substituted at *the Y site* in $YBa_2Cu_3O_{7-\delta}$ cuprate at low doping levels. (In the literature, all Li doping studies at the Cu site were done by more than 5% amount of Li addition.) Despite the extensive reports on the superconducting properties of YBCO doped with different elements, no systematic study had been reported on Li doping at the Y site, to the best of the authors' knowledge, in the literature, especially regarding the high-quality products prepared through the citrate–gel method in this study, compared to the standard solid-state method. The research question raised was whether partial substitution of Li characterized by a high Debye frequency would have any positive effects on $T_c$ and $J_c$ or the electrical and magnetic properties of Y-123. In the BCS theoretical view, it is expected that the superconducting $T_c$ increases owing to high Debye frequency of the atoms in the compound. To answer this question, fully oxygenated $Y_{1-x}Li_xBa_2Cu_3O_{7-\delta}$ polycrystalline samples were synthesized with different levels of Li dopant using an optimum citrate–gel and heat treatment process. A renewed investigation was also carried out on a non-magnetic Li-doped system that had shown enhancements in $T_c$ and $J_c$ values in a previous study of magnetic element doping [23, 24]. We show although the Li dopant was non-magnetic and was capable of bearing a high phonon frequency, no remarkable increases were observed in $T_c$, thereby refuting the hypothesis that $T_c$ increases with phonon frequency. The results were used to shed light on the mechanism of high-$T_c$ superconductivity in cuprates. Moreover, the optimum conditions of the preparation of samples were identified so that we obtained samples with a grain size on the order of 50 μm. This finding of the optimum conditions of the preparation gives nearly identical values of the critical current density ($J_c$) for samples with different Li doping levels (up to the optimum Li doping level $x = 0.01$). In a word, apart from the chemical doping effect, improved preparation conditions would have an extensive effect on $J_c$ enhancement. The present study discloses a contribution to the issue of technological implications of chemically doped citrate-gel prepared YBCO compound. Finally, the citrate–gel synthesis method was exploited in this study due to the demand for homogeneous powders and granule size control required for electrical and magnetic applications of ceramics.

## 2 Materials and methods

### 2.1 Materials

The reagents of $LiNO_3$ (99.99%), $Y(NO_3)_3.6H_2O$ (99.99%), $Ba(NO_3)_2$ (99.99%), and $Cu(NO_3)_2.3H_2O$ (99.99%) powders were purchased from Merck. Deionized water, citric acid ($C_6H_8O_7$), and ethylenediamine ($C_2H_4(NH_2)_2$) were used in the synthesis process.

### 2.2 Citrate-gel synthesis method

Fully oxygenated polycrystalline samples of $Y_{1-x}Li_xBa_2Cu_3O_{7-\delta}$ ($x = 0$, 0.01, 0.02, and 0.1) were synthesized using the citrate–gel method. In citrate–gel method, it could be obtained more homogeneous, single-phased samples and granule size control. For a typical sample, stoichiometric proportions of metal nitrates (0.5 M aqueous solutions) were mixed to obtain a light blue solution to which citric acid was added dropwise at a rate of one drop per 5 s. Next, the pH was raised to 6.8 by adding ethylenediamine dropwise. The final gel was obtained by the constant stirring of the solution at 85 °C for 6 h.

The citrate–gel method involves the formation of a stable complex of cation nitrates in a solvent of water and citric acid. Citric acid takes part in the reaction with a valence of +1 to generate cit-$H_2$ ions, whereas it forms a weak bond with Ba$^{2+}$. Ethylenediamine was, therefore, used as a complexing agent to form ethylenediamine nitrate that prevents Ba$^{2+}$ precipitation. The final gel was heated at 520 °C to yield a dark brown powder via pyrolysis. The resulting powder was ground and heated twice at 880 °C with intermediate grinding. Eventually, the final powder was thoroughly ground again and pressed into rectangular pellets (1 × 3 × 10 mm$^3$) under a 40 bar load before being subjected to the final heating process at 900 °C for 24 h in a pure oxygen flow in order to obtain fully oxygenated samples [25] that were slowly cooled to room temperature. All the samples were prepared concurrently under the same conditions to ensure meaningful and consistent comparisons. The obtained samples show the longest grain size to date; see Fig. 4.

## 2.3 Characterization

The samples were characterized with respect to their crystal structure, phase purity, and lattice parameters via XRD investigations by Cu–K$_\alpha$ radiation using a Philips X'pert MPD system operated at a 2θ range of 10°–90°. The transport $J_c$ data were obtained using a four-probe technique at 77 K. A Philips XL 30 SEM was used to prepare SEM images of the samples. AC magnetic susceptibility measurements were carried out using a Lakeshore Model 7000 susceptometer in a magnetic field of 0.8 A.m$^{-1}$ at a frequency of 333 Hz over a temperature range of 77–100 K. Resistivity was measured using the standard DC four-probe technique. The XRD patterns were analyzed in both X'Pert Highscore 3.0.5 and FullProf Suite version 7.30. The SEM images were studied using Digimizer v. 5.3.5. Also, OriginPro 2019b 9.6.5.169 and VESTA v. 3.4.8 (2019) software tools were used for generating graphic illustrations.

## 3 Results and discussion

### 3.1 Structural analysis by XRD

Figure 1a shows a typical layered perovskite-like the structure of YBCO compound with two CuO$_2$ planes separated by a layer of Y atoms bound by the Ba–O and Cu–O layers. Clearly, the O(2) and O(3) oxygen atoms are strongly coupled with Cu(2) in the CuO$_2$ planes. The XRD patterns were analyzed using the Rietveld refinement method (Fig. 1b). The normalized XRD patterns for pure and Li-doped samples with $x$ = 0.01 and 0.1 are shown in Fig. 1c.

The analyses revealed that the samples were single-phased with an orthorhombic crystal structure of identical perovskite types with $Pmmm$ symmetry (JCPDS 85–1877 reference pattern). The lattice parameters for the pure sample were essentially consistent with those reported in the literature. The X-ray analysis showed identical average oxygen

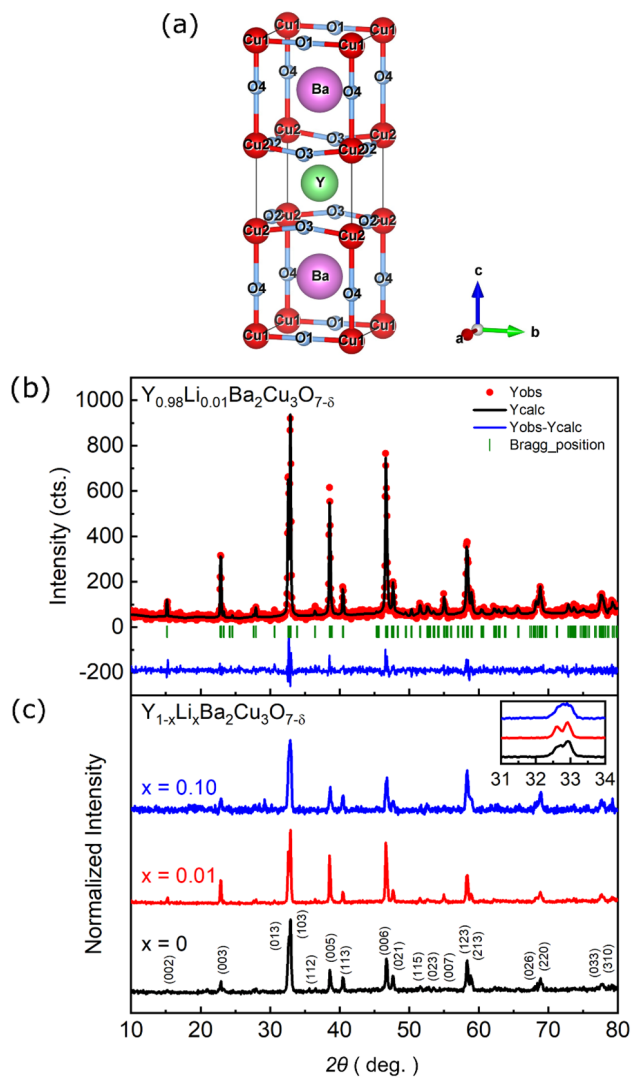

**Fig. 1** **a** Shematic of YBa$_2$Cu$_3$O$_{7-\delta}$ crystal structure, **b** Rietveld refinement plot of powder XRD pattern for Y$_{0.99}$Li$_{0.01}$Ba$_2$Cu$_3$O$_{7-\delta}$ sample, and **c** normalized XRD patterns of YBa$_2$Cu$_3$O$_{7-\delta}$, Y$_{0.99}$Li$_{0.01}$Ba$_2$Cu$_3$O$_{7-\delta}$ and Y$_{0.9}$Li$_{0.1}$Ba$_2$Cu$_3$O$_{7-\delta}$ samples

contents of the samples (7–δ) with a constant value of 6.97 ± 0.02 independent of the substitution degrees [26, 27].

A careful examination in Fig. 1c and comparison of the XRD patterns of both the pure and doped samples reveal strong peaks at (002), (003), (005), (006), and (007), especially for the doped sample. The increasing relative peak intensity in the Li-doped sample indicates grains with a $c$-axis preferred orientation [28].

Clearly, the ionic radii of Li$^+$ at four-, six-, and eightfold coordinated sites are 0.59, 0.76, and 0.92 Å, respectively, while those of Y$^{3+}$ at six- and eightfold coordinated sites are 0.9 and 1.019 Å, respectively. This is while the ionic radii of Cu$^{2+}$ at four- and sixfold coordinated sites are 0.57 and 0.73 Å, respectively, but 0.54 Å for that of Cu$^+$ in the sixfold coordinated sites [29]. Because of the similarity of the ionic radii of $Y^{3+}$ and $Li^+$ at eightfold coordinated sites, at which $Li^+$ is by 9.7% smaller than $Y^{3+}$, and given the fact that $Li^+$ is by 3.5% (4.1%) larger than $Cu^{2+}$ at a 4(6)-fold coordinated site, $Li^+$ ion is seen to move into the $Y^{3+}$ site at low doping levels (as in the present study) [19]. This substitution not only leads to a distortion of $CuO_2$ planes but might also strengthen the two-dimensional character of the charge carriers in these planes [19]. According to Table 1, the $CuO_2$ planes are more buckled by 1% $Li^+$ doping, while the Cu(2)-O(2)-Cu(2) or Cu(2)-O(3)-Cu(2) angles in the $CuO_2$ planes have decreased. This led to an increase in $T_c$, which might be related to orthorhombicity that is known to increase with increasing distance between the (013) and the (103) peaks (see the inset of Fig. 1c). Moreover, because of the $Li^+$ and $Cu^{2+}$ similarity (i.e., 0.73 Å vs 0.76 Å in an octahedral coordination and 0.57 Å vs 0.59 Å in a tetrahedral coordination), $Li^+$ is reportedly able to enter the copper site in the $CuO_2$ planes [21] (at higher doping levels), reduce hole concentration, and cause a $T_c$ suppression as well. In the present study, both these effects had a synergic impact as a result of adding 1% Li that led to a slight reduction in $T_c$ or to maintain a nearly constant $T_c$ value (see Fig. 2). (It may be noted that in the electrical resistivity measurements, $T_c$ experienced a slight increase by 1% Li doping before it started to decrease at higher doping levels; Fig. 3.) It may be claimed that Li moves mainly to the Y site, which may increase $T_c$ due to the strengthened two-dimensional character of the carriers

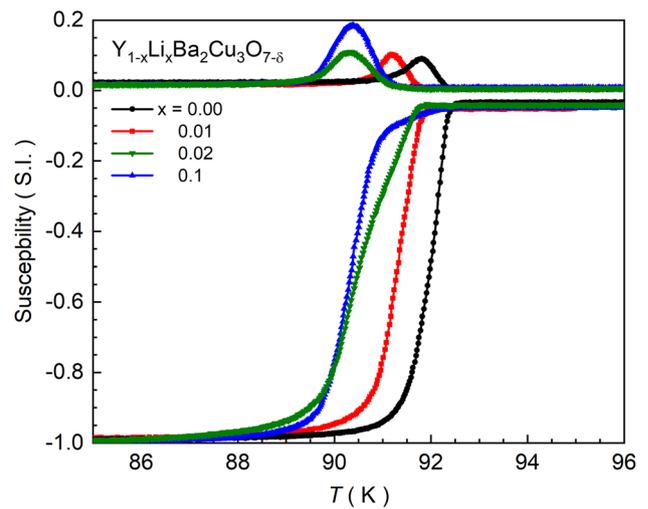

**Fig. 2** AC magnetic susceptibility vs temperature of $Y_{1-x}Li_xBa_2Cu_3O_{7-\delta}$ samples with $x = 0.0$, 0.01, 0.02, 0.1. $T_{conset}$ is defined as the temperature below which the real part of AC susceptibility deviates from its normal-state linear behavior and drops to a diamagnetic superconducting phase

of $CuO_2$ planes [19], but that it is partially substituted at the $Cu^{2+}$ site, which would drastically decrease $T_c$; hence, the overall result is a slight reduction in $T_c$.

Finally, if the $Li^+$ dopant goes into the Cu(1) site of the chains, it would result in a lattice expansion along the $a$-axis, causing a great decrease in $T_c$ [15–17, 19–21]; this was not, however, the case observed in the present study (see Table 1). It should be noted that the addition of impurities to high-$T_c$ cuprates gives rise to variations in $T_c$ mainly due to the related changes in carrier density and oxygen content.

**Table 1** Unit cell parameters, bonding intervals and angles in $Y_{1-x}Li_xBa_2Cu_3O_{7-\delta}$ samples for $x = 0$, 0.01, and 0.1 as obtained by XRD Rietveld refinement. $T_{conset}$ (K) and $T_{c\,mid}$ (K) are experimental superconducting transition temperature obtained from susceptibility measurements

| Li-doping level | 0% | 1% | 10% |
|---|---|---|---|
| Lattice constants (Å) | | | |
| $a$ | 3.8139 | 3.8130 | 3.8183 |
| $b$ | 3.8836 | 3.8783 | 3.8827 |
| $c$ | 11.6643 | 11.6762 | 11.6583 |
| Unit cell volume (Å³) | | | |
| V | 172.7680 | 172.6698 | 172.8398 |
| Bond lengths (Å) | | | |
| Cu(2)–O(2) | 1.9208 (3) | 1.9237 (3) | 1.9564 (5) |
| Cu(2)–O(3) | 1.9683 (4) | 2.0076 (6) | 1.9524 (3) |
| Cu(2)–O(4) | 2.1745 (18) | 2.1846 (18) | 2.3537 (18) |
| Cu(1)–O(4) | 1.9641 (16) | 1.9861 (16) | 1.7758 (16) |
| Cu(2)–Cu(2) | 3.3830 (19) | 3.3295 (19) | 3.4244 (19) |
| Bond angles (deg.) | | | |
| Cu(2)–O(2)–Cu(2) | 166.75 (12) | 165.10 (12) | 155.04 (12) |
| Cu(2)–O(3)–Cu(2) | 161.63 (12) | 150.66 (12) | 168.59 (13) |
| $T_{conset}$ (K) | 92.61 | 92.01 | 92.34 |
| $T_{cmid}$ (K) | 92 | 91.5 | 90.4 |

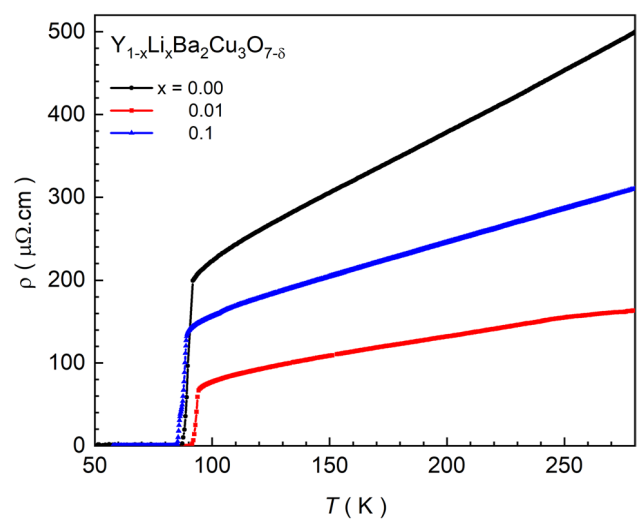

**Fig. 3** The electrical resistivity vs temperature of $Y_{1-x}Li_xBa_2Cu_3O_{7-\delta}$ samples with $x = 0.0$, 0.01, 0.1

This does not, however, mean that the effects of impurity on pair breaking and scattering could be ignored [30].

Table 1 presents the lattice parameters, bonding intervals, and bond angles for pure YBa$_2$Cu$_3$O$_{7-\delta}$ and Li-doped samples. Clearly, the Cu(2)–O(4) distance increased with Li doping, as did the lattice constant $c$. However, Cu(2)–Cu(2) distance (defined as the spacing between CuO$_2$ planes) has been reported to decrease at low doping levels. According to Anderson's theoretical model [26], $T_c$ is predicted to increase with decreasing separation between the CuO$_2$ planes. The same trend has also been reported for LaBaCuO and LaSr-CuO, for which a rapid increase has been observed in $T_c$ as a result of either applying an external pressure or substituting a smaller cation for the La site that reduce the distance between CuO$_2$ planes. This is, however, a controversial point on which there is no universal agreement. Hardy et al. [31], for instance, applied uniaxial pressure along the tetragonal $a$- and $c$-axes for an optimally doped HgBa$_2$CuO$_{4+\delta}$ (Hg–1201) cuprate. By bringing the CuO$_2$ planes closer together through the pressure applied along the $c$-axis, $T_c$ was expected to decrease rapidly in accordance with Anderson's theoretical model, while it was observed to increase as the area of the CuO$_2$ plane declined [26]. Also, Locquet et al. observed an almost twofold increase in $T_c$ in LaSrCuO$_4$ thin films epitaxially grown on a suitable substrate, which would increase the separation between the CuO$_2$ planes [32]. In a recent review, Schilling et al. [33] concluded that $T_c$ in optimally doped cuprates could increase in two ways: by increasing the separation between CuO$_2$ planes or by reducing their plane area. The dependence of $T_c$ on the separation between planes or plane area differs widely from one experimental group to another in Fe-based superconductors that form another class of nearly 2D-layered superconductors in which metallic layers of Fe atoms form a planar square lattice bonded with as atoms above and below them [33]. For instance, by substituting smaller rare-earth cations for La in LaFeAsO$_{1-x}$F$_x$ (La–1111) and reducing the separation between the conduction FeAs layers, $T_c$ exhibits a great increase [34]. In Y-doped SmFeAsO compounds [35], however, both enhancements and reductions in $T_c$ have been reported despite the reductions shown by the compound in its $a$ and $c$ lattice parameters.

## 3.2 AC magnetic susceptibility results

An increase in $T_c$ is expected to occur by Li addition due to its high Debye frequency and light atomic weight. As elaborated in the BSC theory, higher Debye frequencies would lead to higher $T_c$ values. The AC magnetic susceptibility results exhibit the $T_c$ values reported in Fig. 2. In this case, $T_{conset}$ is defined as the temperature below which the real part of AC susceptibility deviates from its normal-state linear behavior and drops to a diamagnetic superconducting phase. As illustrated in Fig. 2, the superconducting transition is sharp for pure and Li$_{0.01}$-doped samples with no secondary steps on the transition curve, showing the samples are homogeneous and single-phased. Moreover, the transition temperatures of the doped samples are only slightly lower than that of the pure sample; for pure and 10% doped samples, $T_{c\,mid}$ is 92 K and 90.4 K, respectively. (see Table 1). In other words, $T_c$ is almost unchanging even for a Li doping of 10%. This is consistent with the magnetic coupling models and observations reported for cuprate superconductors [23, 24], where by magnetic element doping at the Y site (Co, Fe and Ni element doping) $T_c$ increased, while the substitution of two non-magnetic elements, i.e., Ca and Sr, results in the $T_c$ degradation. The increase in $T_c$ through magnetic doping, which was a new observation, confirms the effective role of magnetically mediated pairing in cuprates [24]. However, a remarkable reduction in $T_c$ has been observed for a substitution of Li for Cu that is attributed to the lower hole concentration [29]. To note, comparing with the recent observations in cuprate superconductors [24], where the doped sample with $T_c$ nearly to the pure one is an optimum doped sample, here, we could consider $x=0.01$ as an optimum doping level of Li. Therefore, we focused mainly on pure, $x=0.01$ and a higher doping level of Li, such as $x=0.1$, in most of the analyses.

## 3.3 Electrical resistivity

Figure 3 exhibits the temperature-dependent resistivity of Y$_{1-x}$Li$_x$Ba$_2$Cu$_3$O$_{7-\delta}$ samples with $x=0$, 0.01, and 0.1 measured in a zero magnetic field. In the normal state down to above $T_c$, the temperature dependence is linear for all the prepared samples, exhibiting a transition from a metallic behavior to a superconducting one. The linear behavior of $\rho(T)$, a distinctive feature observed at optimal doping levels in high-$T_c$ cuprates, shows that all the samples are fully oxygenated at the optimal carrier concentration [24, 36] such that $T_c$ remains nearly constant when 10% Li is added.

## 3.4 SEM analysis

Figure 4 presents the SEM micrographs of the samples at resolutions of 10 μm (Left) and 20 μm (Right). These polycrystalline samples show the longest grain size to date when compared with the microstructural analyses reported so far on the different bulk preparations of cuprates. These observations bear witness to the optimum preparation conditions employed in the present study. Grains as large as 50 μm are found in the samples along with closely packed smaller ones (Fig. 4). The large and closely packed grains contribute to the higher $J_c$ values of the samples in comparison with the values reported for similar samples prepared using the solid-state method (Fig. 5). The pure sample exhibits a $J_c$ of around 100 A.cm$^{-2}$ [23] for the solid-state sample, whereas a value of 7 times higher was recorded for the citrate–gel samples.

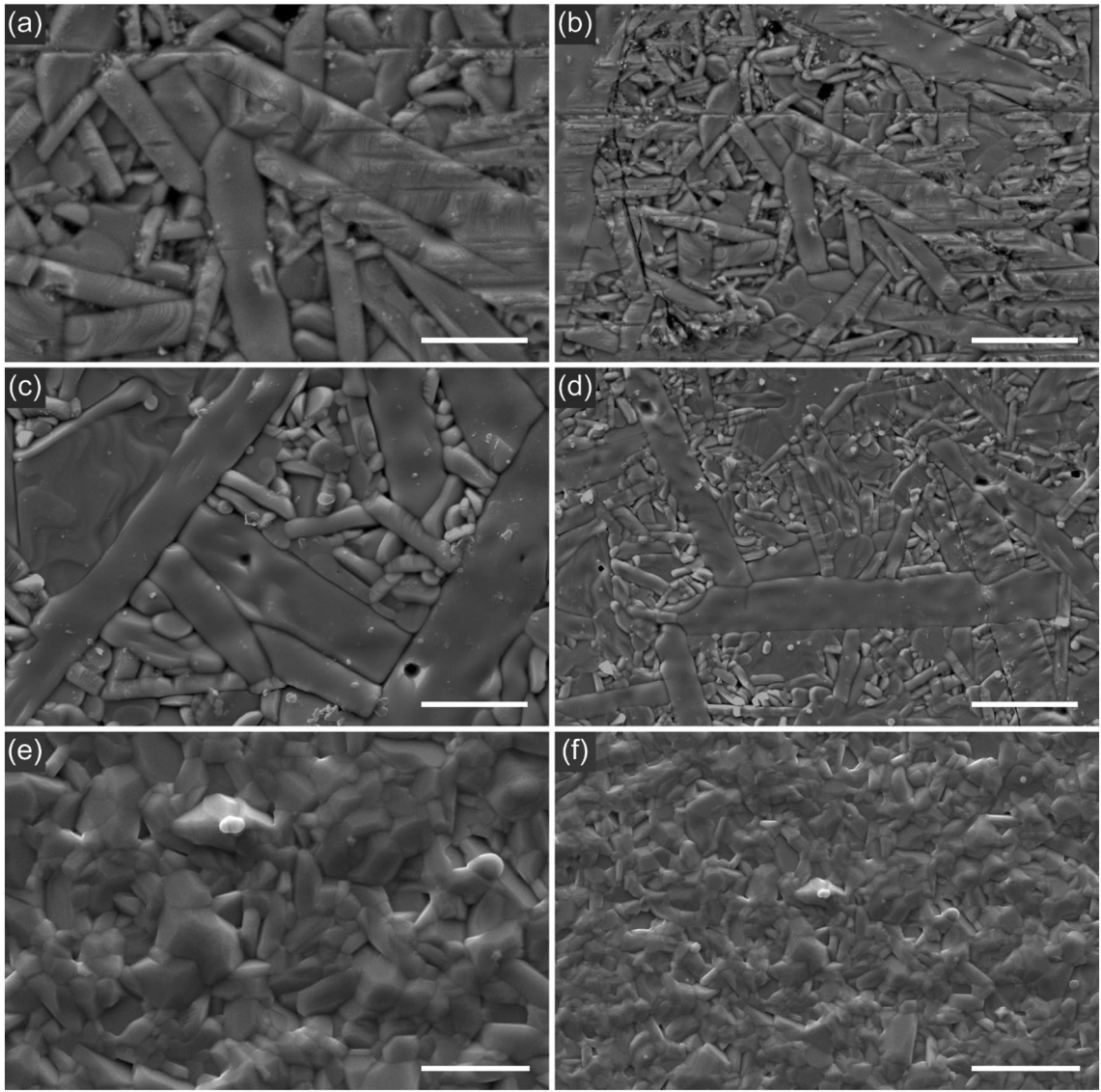

**Fig. 4** SEM micrographs (scanning electron microscope) of $Y_{1-x}Li_xBa_2Cu_3O_{7-\delta}$ samples with **a, b** $x = 0$, **c, d** $x = 0.01$ and **e, f** $x = 0.1$. The micrographs are shown in 10 μm (Left) and 20 μm (Right) resolutions

### 3.5 $J_c$ measurements

The V–J characteristic curves for the $Y_{1-x}Li_xBa_2Cu_3O_{7-\delta}$ samples ($x = 0$, 0.01, 0.1) measured at 77 K are also presented in Fig. 5. The $J_c$ value is determined by extrapolating the linear part of the V–J curve at the intersection with the horizontal current density ($J$) axis. The maximum $J_c$ values obtained for the $x = 0$ and $x = 0.01$ samples were 680 and 633 A.cm$^{-2}$, respectively. The 1% Li doping is the concentration with the least $T_c$ reduction (in the corresponding magnetic susceptibility measurements) and maximum $J_c$ (i.e., optimum concentration); higher concentrations of Li above the optimum value led to a lower pinning strength and reduced $J_c$ (see Fig. 5). By introducing the optimum Li dopant, no remarkable change was detected in the magnitude of $J_c$, indicating that the sample preparation and heating process was performed under optimum conditions; hence, no considerable differences or effects were observed either in the pinning properties or in $J_c$ values as a result of adding Li. However, some studies have indicated that $J_c$ would exhibit a more than twofold increase in Li-substituted YBCO prepared by the melt-textured method when compared with the plain undoped compound [37]. Adding impurity normally affects the pinning properties in cases where partial

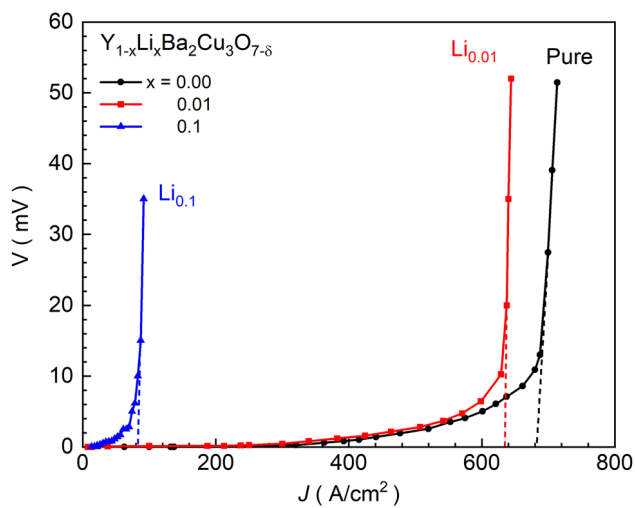

**Fig. 5** V–J characteristic curves of $Y_{1-x}Li_xBa_2Cu_3O_{7-\delta}$ samples ($x = 0$, 0.01 and 0.1) at 77 K. $J_c$ of each sample is where the extrapolation of the linear part of the V–J curve, cut the horizontal axis. By increasing Li dopant, there is no big change in the value of $J_c$

clustering of the dopant or dislocations appears in the crystal structure.

Despite the remarkable advances made in achieving high critical current densities in cuprates by neutron, electron, or heavy ion irradiation and in spite of the extensive research conducted to determine the effects of chemical substitutions, oxygen deficiency, or structural defects on the pinning mechanism in cuprates, the main problem of increasing flux pinning due to chemical doping in YBCO yet remains to be resolved. It has been found that chemical substitutions into $CuO_2$ planes and/or partial clustering of the dopant between the grains form effective pinning centers in the $YBa_2Cu_3O_{7-\delta}$ compound. The main disadvantage of the conventional preparation method is the need for a large number of pinning centers to increase $J_c$, which could lead to a drastic $T_c$ suppression. The improved preparation process proposed in the current study, however, led to a profound effect on $J_c$ enhancement, with nearly no $T_c$ suppression. High critical current density is crucial for the technological applications of bulk high-temperature superconductors.

### 3.6 Prediction: machine learning results

The effect of Li doping on the superconducting temperature of $Y_{1-x}Li_xBa_2Cu_3O_{7-\delta}$ samples was studied by machine learning method, too.

Physicists attempt to develop mathematically based models that can predict the future behavior of a system as well as effectively predict the outcomes of experiments that have not yet been measured [38]. Machine learning methods are currently one of the most effective approaches for analyzing data and extracting a suitable model from the dataset for predicting the results of new experiments [39]. In recent years, the availability of huge data and the improvement of machine learning-based algorithms have caused an unprecedented increase in the use of machine learning methods [40–44].

Algorithms and datasets are two essential tools for research in data science. In this work, the CatBoost algorithm was used to predict the transition temperature of $Y_{1-x}Li_xBa_2Cu_3O_{7-\delta}$ cuprate compound and to compare it with the presented experimental results. The CatBoost algorithm is a method based on gradient-boosted decision trees (GBDT) machine learning ensemble techniques. GBDT are an effective tool for solving regression and classification problems in big data [45]. Here, utilizing the CatBoost algorithm and tuning its hyperparameters were done for the first time in this field of research.

To predict the transition temperature of superconducting materials, an algorithm must be trained with a dataset. The SuperCon is now the world's largest and most complete superconducting materials database, which is utilized in the field of methods based on artificial intelligence [46]. After the data preprocessing steps, 16,148 superconducting compounds were considered a dataset. 90% of the dataset was utilized for training the algorithm, while 10% of the dataset, i.e., 1614 compound, was reserved for the testing phase.

The ID and characteristics of each superconducting compound are the features that we make for the compounds; therefore, each compound can be known to the algorithm by its features. Here, 132 features were obtained for each compound using the Magpie package, and then, 25 of the most important features were selected utilizing the CatBoost algorithm's capability [47]. Then, considering that the attributes play a crucial role in predicting the transition temperature, 241 other features were generated for each superconducting compound, and among them, 30 of the most important features were selected using the CatBoost algorithm's capability.

In addition, two attributes were added to complete the set of features: the number of elements in each composition and the sum of each composition's subscript.

Depending on the number of various tasks performed on the dataset and choosing the suitable algorithm, the prediction accuracy of developed models based on machine learning differs from one another. In order to compare the accuracy and efficiency of the developed models, there are evaluation criteria, the two most important of which are: R-squared ($R^2$) and root mean square error (RMSE) [38]:

$$R^2 = 1 - \frac{\sum_{i=1}^{m}(y_i - \hat{y}_i)^2}{\sum_{i=1}^{m}(y_i - \bar{y})^2} \quad RMSE = \sqrt{\frac{1}{m}\sum_{i=1}^{m}(y_i - \hat{y}_i)^2}$$

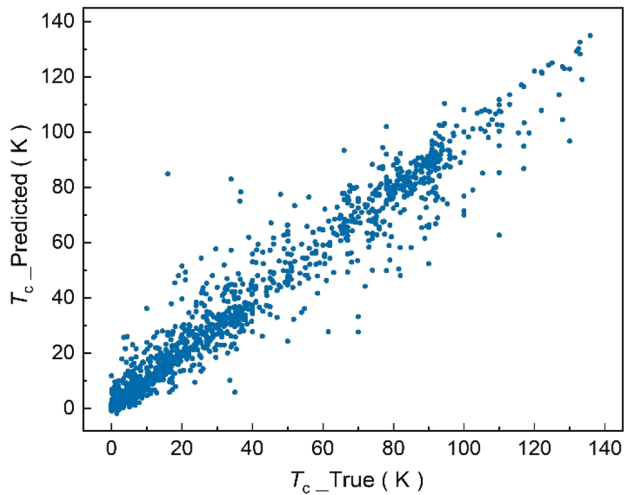

**Fig. 6** Superconducting transition temperature predicted by the model and compared to the true values (for 1614 superconducting compounds)

**Table 2** The transition temperatures predicted by the model for $Y_{1-x}Li_xBa_2Cu_3O_7$ cuprate

| Compound | $T_c$_prediction(K) |
|---|---|
| $YBa_2Cu_3O_7$ | 92.46 |
| $Y_{0.99}Li_{0.01}Ba_2Cu_3O_7$ | 91.25 |
| $Y_{0.98}Li_{0.02}Ba_2Cu_3O_7$ | 90.11 |
| $Y_{0.9}Li_{0.1}Ba_2Cu3O_7$ | 83.86 |

where $m$ represents the number of superconducting compounds, and $y_i$ indicates the true transition temperature of the $i$-th sample which exists in the dataset. $\hat{y}_i$ represents the transition temperature of the $i$-th sample that predicted by the model, and $\bar{y}$ is the average value corresponding to the true values of $m$ samples. In this study, $R^2$ value of 0.949 and RMSE value of 7.59 Kelvin were obtained, which are suitable values comparing with similar studies [47–52], indicating that the model developed in this study is reliable.

Using the model developed in this study, the transition temperature for ten percent of the data set, or 1614 superconducting compounds, is predicted and compared to the true values in Fig. 6. Each point in Fig. 6 represents a superconducting compound's transition temperature.

Finally, the CatBoost algorithm performed the prediction process using 57 features for each compound. The transition temperatures predicted by the model for $Y_{1-x}Li_xBa_2Cu_3O_{7-\delta}$ cuprate compounds are shown in Table 2. It can be seen that $T_c$ is nearly decreasing with Li doping.

More innovations are associated with the model, including:

- using novel features for superconducting materials that were not utilized in prior studies. Some of these features include electron affinity, Pettifor number, magnetic moment, dipole polarizability, thermal conductivity, etc.
- using various of feature selection methods to pick the most important features for model development.
- applying many data preprocessing steps to the dataset.

## 4 Conclusion

The effects of non-magnetic Li doping at the Y site in $Y_{1-x}Li_xBa_2Cu_3O_{7-\delta}$ ($x = 0$, 0.01, 0.02, 0.1) on the $T_c$ and $J_c$ values as well as the crystal structure of the compound were investigated. The optimum conditions for the citrate-gel preparation method were determined. The surface of the samples thus prepared was examined by SEM to find that the grains were closely packed with no porosity, while the size of the largest grain reached 50 μm. Moreover, no variation was observed in $J_c$ values as a result of the optimum Li doping. Rietveld refinement analysis of XRD data showed that the $c$ lattice parameter increased with Li doping, while the distance between $CuO_2$ planes decreased. AC magnetic susceptibility bulk measurements showed that although the Li dopant was non-magnetic and was capable of bearing a high phonon frequency, no remarkable increases were observed in $T_c$, thereby refuting the hypothesis that $T_c$ increases with phonon frequency. This is consistent with magnetic coupling models and recent observations in cuprate superconductors. The CatBoost algorithm was also used to predict the transition temperature of $Y_{1-x}Li_xBa_2Cu_3O_{7-\delta}$ cuprates. It showed that $T_c$ is nearly decreasing with Li doping consistent with the experimental results. Finally, it was found that improving upon the preparation process would have a profound effect on $J_c$ enhancement (by a factor of 7 compared with the standard solid-state method) with nearly no $T_c$ suppression, which seems promising for further fundamental and applied research. It would be interesting to investigate the effect of lower doping of Li, $x < 0.01$, on the $T_c$, $J_c$ and pinning force of $Y_{1-x}Li_x$-123. The question is worthy to be asked, and further experiments are planned in this regard.

**Funding** No funding was received for conducting this study.

## Declarations

## References

1.  H.K. Onnes, Nobel Lecture **4**, 306 (1913)

2. J.X. Lin, X.M. Liu, C.W. Cui, C.Y. Bai, Y.M. Lu, F. Fan, Y.Q. Guo, Z.Y. Liu, C.B. Cai, Adv. Manuf. **5**, 165 (2017). https://doi.org/10.1007/s40436-017-0173-x

3. D. Perconte, K. Seurre, V. Humbert, C. Ulysse, A. Sander, J. Trastoy, V. Zatko, F. Godel, P.R. Kidambi, S. Hofmann, X.P. Zhang, D. Bercioux, F.S. Bergeret, B. Dlubak, P. Seneor, J.E. Villegas, Phys. Rev. Lett. **125**, 087002 (2020). https://doi.org/10.1103/PhysRevLett.125.087002

4. A. Ramos-Alvarez, N. Fleischmann, L. Vidas, A. Fernandez-Rodriguez, A. Palau, S. Wall, Phys. Rev. B **100**, 184302 (2019). https://doi.org/10.1103/PhysRevB.100.184302

5. A. Gauzzi, Y. Klein, M. Nisula, M. Karppinen, P.K. Biswas, H. Saadaoui, E. Morenzoni, P. Manuel, D. Khalyavin, M. Marezio, T.H. Geballe, Phys. Rev. B **94**, 180509 (2016). https://doi.org/10.1103/PhysRevB.94.180509

6. J.C. Bernardi, D.A. Modesto, M.S. Medina, A. Zenatti, E.C. Venâncio, E.R. Leite, A.J.C. Lanfredi, M.T. Escote, Mater. Res. Express **6**, 086001 (2019). https://doi.org/10.1088/2053-1591/ab1fa3

7. R.V. Vovk, G. Ya Khadzhai, O.V. Dobrovolskiy, Z.F. Nazyrov, I.L. Goulatis, Mater. Res. Express **1**, 026303 (2014). https://doi.org/10.1088/2053-1591/1/2/026303

8. G. Krabbes, G. Fuchs, P. Schätzle, S. Gruß, J. Park, F. Hardinghaus, G. Stöver, R. Hayn, S.-L. Drechsler, T. Fahr, Physica C **330**, 181 (2000). https://doi.org/10.1016/S0921-4534(99)00616-4

9. L. Shlyk, G. Krabbes, G. Fuchs, Physica C **390**, 325 (2003). https://doi.org/10.1016/S0921-4534(03)00737-8

10. F. Yong, Z. Lian, Physica C **202**, 298 (1992). https://doi.org/10.1016/0921-4534(92)90174-B

11. L. Shlyk, G. Krabbes, G. Fuchs, G. Stöver, S. Gruss, K. Nenkov, Physica C **377**, 437 (2002). https://doi.org/10.1016/S0921-4534(01)01298-9

12. A. Mahajan, H. Alloul, G. Collin, J. Marucco, Eur. Phys. J. B-Condens. Matter Complex Sys. **13**, 457 (2000). https://doi.org/10.1007/s100510050058

13. L. Shlyk, G. Krabbes, G. Fuchs, K. Nenkov, B. Schüpp, J Appl. Phys. **96**, 3371 (2004). https://doi.org/10.1063/1.1778215

14. A. Mahajan, H. Alloul, G. Collin, J. Marucco, Phys. Rev. Lett. **72**, 3100 (1994). https://doi.org/10.1103/PhysRevLett.72.3100

15. F. Maury, M. Nicolas-Francillon, F. Bouree, I. Mirebeau, Physica C **341**, 609 (2000). https://doi.org/10.1016/S0921-4534(00)00612-2

16. M. Nicolas-Francillon, K. Sauv, J. Conard, F. Bouree, Physica C **273**, 49 (1996). https://doi.org/10.1016/S0921-4534(96)00495-9

17. M. Nicolas-Francillon, F. Maury, R. Ollitrault-Fichet, M. Nanot, P. Legeay, J. Appl. Phys. **84**, 925 (1998). https://doi.org/10.1063/1.368157

18. F. Maury, M. Nicolas-Francillon, F. Bourée, R. Ollitrault-Fichet, M. Nanot, Physica C **333**, 121 (2000). https://doi.org/10.1016/S0921-4534(00)00083-6

19. Q. Gao, Y. Zhou, L. Zhang, H. Wang, Physica C **199**, 121 (1992). https://doi.org/10.1016/0921-4534(92)90548-Q

20. P. Mukherjee, A. Simon, M. Sarma, A. Damodaran, Solid State Commun. **81**, 253 (1992). https://doi.org/10.1016/0038-1098(92)90509-8

21. G.H. Kewi, K.C. Ott, E.J. Peterson, Physica C **194**, 307 (1992). https://doi.org/10.1016/S0921-4534(05)80009-7

22. K. Inoue, N. Sakai, M. Murakami, I. Hirabayashi, Physica C **445**, 128 (2006). https://doi.org/10.1016/j.physc.2006.03.093

23. B. Hadi-Sichani, H. Shakeripour, H. Salamati, Mater. Res. Express **7**, 056002 (2020). https://doi.org/10.1088/2053-1591/ab903e

24. H. Shakeripour, S.S. Hosseini, S.S. Ghotb, B. Hadi-Sichani, S. Pourasad, Ceram. Int. **47**, 10635 (2021). https://doi.org/10.1016/j.ceramint.2020.12.176

25. H. Shakeripour, M. Akhavan, Supercond. Sci. Technol. **14**, 234 (2001). https://doi.org/10.1088/0953-2048/14/5/302

26. P.W. Anderson, Science **279**, 1196 (1998). https://doi.org/10.1126/science.279.5354.1196

27. H. Shakeripour, M. Akhavan, Supercond. Sci. Technol. **14**, 213 (2001). https://doi.org/10.1088/0953-2048/14/4/306

28. W. Wang, Q. Chen, Q. Cui, J. Ma, H. Zhang, Physica C **511**, 1 (2015). https://doi.org/10.1016/j.physc.2015.02.003

29. R.D. Shannon, Acta Crystallographica Sect. A **32**, 751 (1976). https://doi.org/10.1107/S0567739476001551

30. A.A. Abrikosov and L.P. Gor'kov, Zhur. Eksptl'. I Teoret. Fiz. **39**, 1781 (1960) [Sov. Phys. JETP **12**, 1243 (1961)]

31. F. Hardy, N. Hillier, C. Meingast, D. Colson, Y. Li, N. Barišić, G. Yu, X. Zhao, M. Greven, J. Schilling, Phys. Rev. Lett. **105**, 167002 (2010). https://doi.org/10.1103/PhysRevLett.105.167002

32. J.-P. Locquet, J. Perret, J. Fompeyrine, E. Mächler, J.W. Seo, G. Van Tendeloo, Nature **394**, 453 (1998). https://doi.org/10.1038/28810

33. J. Schilling, N. Hillier, N. Foroozani, J. Phys. **449**, 012021 (2013). https://doi.org/10.1088/1742-6596/449/1/012021

34. Z.A. Ren, W. Lu, J. Yang, W. Yi, X.L. Shen, Z.C. Li, G.C. Che, X.L. Dong, L.L. Sun, F. Zhou, Chin. Phys. Lett. **25**, 2215 (2008). https://doi.org/10.1088/0256-307X/25/6/080

35. K. Lai, F. Kwong, D.H. Ng, J. Appl. Phys. **111**, 093912 (2012). https://doi.org/10.1063/1.4712309

36. Y. Chen, Y. Cui, C. Cheng, Y. Yang, Y. Zhang, Y. Zhao, J. Supercond. Novel Magn. **23**, 621 (2010). https://doi.org/10.1007/s10948-010-0698-8

37. T. Ito, K. Takenaka, S. Uchida, Phys. Rev. Lett. **70**, 3995 (1993). https://doi.org/10.1103/PhysRevLett.70.3995

38. W. Lo, Y.X. Zhou, T. Tang, K. Salama, Physica C **354**, 152 (2001). https://doi.org/10.1016/S0921-4534(01)00133-2

39. C.J. Stark, Phys. Rev. A **97**, 020103 (2018). https://doi.org/10.1103/PhysRevA.97.020103

40. P. Mehta, M. Bukov, C.-H. Wang, A.G.R. Day, C. Richardson, C.K. Fisher, D.J. Schwab, Phys. Rep. **810**, 1–124 (2019). https://doi.org/10.1016/j.physrep.2019.03.001

41. J. Schmidt, M.R.G. Marques, S. Botti, M.A.L. Marques, npj Comput. Mater. **5**, 1–36 (2019)

42. Y. Zhang, X. Xu, Physica C **595**, 1354031 (2022). https://doi.org/10.1016/j.physc.2022.1354031

43. Y. Zhang, X. Xu, Physica C **597**, 1354062 (2022). https://doi.org/10.1016/j.physc.2022.1354062

44. Y. Zhang, X. Xu, Physica C **592**, 1353998 (2022). https://doi.org/10.1016/j.physc.2021.1353998

45. Y. Zhang, X. Xu, J. Supercond. Novel Magn. **34**, 63–73 (2021). https://doi.org/10.1007/s10948-020-05682-0

46. J.T. Hancock, T.M. Khoshgoftaar, J. Big Data **7**, 1–45 (2020). https://doi.org/10.1186/s40537-020-00369-8

47. B. Roter, S. Dordevic, Physica C **575**, 1353689 (2020). https://doi.org/10.1016/j.physc.2020.1353689

48. Y. Dan, R. Dong, Z. Cao, X. Li, C. Niu, S. Li, J. Hu, IEEE Access **8**, 57868–57878 (2020). https://doi.org/10.1109/ACCESS.2020.2981874

49. K. Hamidieh, Comput. Mater. Sci. **154**, 346–354 (2018). https://doi.org/10.1016/j.commatsci.2018.07.052

50. T. Konno, H. Kurokawa, F. Nabeshima, Y. Sakishita, R. Ogawa, I. Hosako, A. Maeda, Phys. Rev. B **103**, 014509 (2021). https://doi.org/10.1103/PhysRevB.103.014509

51. V. Stanev, C. Oses, A.G. Kusne, E. Rodrigues, J. Paglione, S. Curtarolo, I. Takeuchi, npj Comput Mater **4**, 1–14 (2018). https://doi.org/10.1038/s41524-018-0085-8

52. S. Li, Y. Dan, X. Li, T. Hu, R. Dong, Z. Cao, J. Hu, Symmetry **12**, 262 (2020). https://doi.org/10.3390/sym12020262